\documentclass[letterpaper,10pt,conference]{ieeeconf} 

\IEEEoverridecommandlockouts   
\usepackage{mathtools,amsmath,amssymb}
\usepackage{color,soul,graphicx,etoolbox}
\usepackage{algorithm}
\usepackage{algpseudocode}
\usepackage{comment}
\usepackage{tikz}
\usepackage{trimclip}
\usepackage{float}
\DeclareMathOperator*{\argmin}{arg\,min}

\newcommand{\ParFor}[1]{\For{ #1 }\textbf{ in parallel}}
\newtheorem{remark}{Remark}

\newcommand{\red}[1]{\textcolor{red}{#1}}

\newcommand{\norm}[1]{\left\lVert#1\right\rVert}
\newcommand{\diag}[2]{\begin{bmatrix}#1 & & \\ & \ddots & \\ &  & #2\end{bmatrix}}
\newcommand{\ColVec}[2]{\begin{bmatrix} #1 \\ \vdots \\ #2\end{bmatrix}}

\newcommand{\mat}[1]{\begin{bmatrix}#1\end{bmatrix}}
\newcommand{\costFcn}[3]{\min_{#1} ~~ & #2 \\ \text{s.t.} ~~ & #3}
\newcommand{\sddots}{\begin{trimbox}{0pt 0pt 0pt 5pt}{$\ddots$}\end{trimbox}}
\newcommand{\bz}{\mathbf{z}}

\tikzset{
	pics/valve/.style args={#1,#2}{
		code={
			\draw [very thick] (0,-0.25) -- (1,0.25) -- (1,-0.25) -- (0,0.25) -- (0,-0.25);
			\node[] (#1_in) at (0.15,0) {};
			\node[] (#1_out) at (0.85,0) {};
			\node[] () at (0.5,0.5) {#2};
		}
	},
	pics/tank/.style args={#1,#2}{
		code={
			\draw [very thick] (0,-0.25) rectangle (1, 0.25);
			\draw [very thick] (0, 0.25) -- (0, 0.75);
			\draw [very thick] (1, 0.25) -- (1, 0.75);
			\node[] (#1_left) at (0.145,0) {};
			\node[] (#1_right) at (0.85,0) {};
			\node[] (#1_bottom) at (0.5,-0.1) {};
			\node[] (#1_top) at (0.5,0.15) {};
			\node[] () at (0.5,0.8) {#2};
		}
	},
	pics/pump/.style args={#1,#2}{
		code={
			\draw [very thick] (0.5,0) circle [radius = 0.5];
			\draw [very thick] (1, 0) -- (0.16, 0.3536) -- (0.16, -0.3536) -- (1, 0);
			\node[] (#1_in) at (0.15,0) {};
			\node[] (#1_out) at (0.85,0) {};
			\node[] () at (0.5,0.8) {#2};
		}
	},
	pics/pumpVert/.style args={#1,#2}{
		code={
			\draw [very thick] (0,0.5) circle [radius = 0.5];
			\draw [very thick] (0, 1) -- (0.3536, 0.16) -- ( -0.3536, 0.16) -- (0, 1);
			\node[] (#1_in) at (0,0.15) {};
			\node[] (#1_out) at (0,0.85) {};
			\node[] () at (0.9,0.5) {#2};
		}
	},
}

\title{\huge\bf Parallel Shooting Sequential Quadratic Programming for Nonlinear MPC Problems\thanks{This research was performed within the framework of the research program AquaConnect, funded by the Dutch Research Council (NWO, grant-ID P19-45) and public and private partners of the AquaConnect consortium and coordinated by Wageningen University and Research.}}

\author{P. C. N. Verheijen$^{1}$, M. Haghi$^{1}$, M. Lazar$^{1}$ and D. Goswami$^{1}$
\thanks{$^{1}$Department of Electrical Engineering, Eindhoven University of Technology, The Netherlands: 
{\tt\small p.c.n.verheijen@tue.nl, m.lazar@tue.nl, d.goswami@tue.nl}}
} 

\begin{document}
\maketitle

\begin{abstract}%
In this paper, we propose a parallel shooting algorithm for solving nonlinear model predictive control problems using sequential quadratic programming. This algorithm is built on a two-phase approach where we first test and assess sequential convergence over many initial trajectories in parallel. However, if none converge, the algorithm starts varying the Newton step size in parallel instead. Through this parallel shooting approach, it is expected that the number of iterations to converge to an optimal solution can be decreased. Furthermore, the algorithm can be further expanded and accelerated by implementing it on GPUs. We illustrate the effectiveness of the proposed Parallel Shooting Sequential Quadratic Programming (PS-SQP) method in some benchmark examples for nonlinear model predictive control. The developed PS-SQP parallel solver converges faster on average and especially when significant nonlinear behaviour is excited in the NMPC horizon. 
\end{abstract}

\section{Introduction}
Model Predictive Control (MPC) is a control strategy that computes the control input by optimizing the predicted response over a finite time horizon while respecting system constraints \cite{RawlingsBook}. Due to this explicit form of computing the control law, MPC problems are significantly more complex compared to classical frequency domain or state feedback controllers. However, the MPC framework is not limited to LTI systems and can be extended to control nonlinear systems. 

There is a rise in applications of nonlinear MPC (NMPC) to high-tech fast systems (e.g., motion control, autonomous vehicles, robotics, power electronics) and to large-scale interconnected systems (micro-grids, water networks, heating/cooling networks). In most of these applications real-time implementation of NMPC is hampered by efficiently and reliably solving the corresponding constrained nonlinear program (NLP). 

This generated much interest in the development of reliable solvers for NMPC, out of which Sequential Quadratic Programming (SQP) methods \cite{NumOptBook:Nocedal}\cite{SQP:Gros2016} or interior point methods \cite{IPOPT:Wachter2006} or projected gradient methods \cite{cGMRES:Ohtsuka2004} have become very popular. SQP in particular is a very effective approach due to its simplicity (successive linearization and solving QPs) and the many reliable QP solvers (qpOases \cite{qpOASES:Ferreau2014}, OSQP \cite{OSQP:Stellato2020}, Hildreth, etc.). It is therefore also widely implemented in available off-the-shelf NLP solvers (e.g., NLPQL(P) \cite{NLPQL:Schittkowski}, SNOPT \cite{SNOPT:Gill2005}, or KNITRO \cite{NumOptBook:Nocedal}). Currently, the main challenges with SQP lie with non-positive definite Hessians, choosing the step size and limiting the number of iterations \cite{NumOptBook:Nocedal}, which implicitly reduces the number of QPs to be solved. Non-monotone step-size selecting algorithms, for example in \cite{NewtonNL:Grippo1986}\cite{NLPQL:Schittkowski}, guarantee convergence, but often at the cost of the number of iterations. Similarly, Quasi-Newton methods ensure a positive--definite Hessian, but lose the benefits of using sparse Hessians, such as efficient numerical calculations and a low memory profile. 

Besides the advances in solvers for nonlinear programming, efforts to parallelize NMPC algorithms or numerical solvers have been made, which coupled with parallel computing architectures (multi-core, FPGA and GPU) can speed-up NMPC algorithms. In the literature, the parallelization strategy can be subdivided into 2 categories. At one end, shooting methods, such as using Monte--Carlo estimates \cite{ParallelNMPConGPU:Ohyama2017}, Particle-Swarm-Optimization \cite{NMPConFPGA:Fang2021}, policy iteration with parallel computing in each predicted time sample (but sequential over the horizon) \cite{TowardsParNMPC:Bobiti2017} or solving many QPs in parallel for different initial guesses \cite{GPUaccSQP:Hu2017}. On the other end, operator splitting or distributed methods, including Newton methods with a correction step \cite{NewtonParallel:Deng2018}\cite{ALADINParSQP:Kouzoupis2016}, alternating direction method of multipliers \cite{GPUADMM:Schubiger2020} or accelerated proximal gradient methods \cite{GPU_SMPC_WDN:Sampathirao2018} have been developed.


Motivated by this state-of-the-art, we propose two approaches based on parallel shooting for accelerating SQP in general, and SQP for NMPC in particular. The first approach is to explore multiple initial trajectories in parallel. The second approach shoots different step-sizes and tests these in parallel, where each sequential step is linearized based on the results of the previous step with the fastest convergence. We show that the first approach promotes fast converging trajectories, while the second approach prevents iterations from getting stuck in a cycle. Therefore, by combining these approaches into two phases, the proposed PS-SQP algorithm accelerates convergence with respect to classical SQP. A further advantage of the parallel shooting approach to SQP is that it lends itself very well to implementation on GPUs, which will be the subject of future research.
It is worth to mention that a parallel SQP implementation for GPUs was already presented in \cite{GPUaccSQP:Hu2017}, which proposed to solve multiple QPs in parallel within a GPU architecture. However, therein, no detail about the initialization of multiple QPs was given.

Next, the preliminary information on SQP is presented in Section \ref{sec:Prelim}. In Section \ref{sec:PSSQP} we discuss and explain the parallel implementation of our approach on SQP. The proposed algorithm is illustrated on 2 systems in Section \ref{sec:Examples}. First, an inverted pendulum is considered to illustrate the benefits on highly nonlinear systems. Secondly, a small water network is considered to show the applicability of PS-SQP for large scale NMPC problems. The conclusions are summarized in Section \ref{sec:Conclusion}.

\section{Sequential Quadratic Programming for NMPC}\label{sec:Prelim}
Sequential Quadratic Programming (SQP) solves a nonlinear problem by sequentially linearizing the problem over its current operating point. The operating point (that we shall further on refer to as a "trajectory") is updated with the optimal solution from the linearized Quadratic Program (QP). Although multiple variations of sequential quadratic programming exist, we will adopt in what follows the version from \cite{SQP:Gros2016}. Consider the following nonlinear MPC problem:
\begin{equation}
    \label{eq:NMPC_problem}
    \begin{aligned}
        \min_{x_{i|k}, u_{i|k}} ~~ &\sum_{i=0}^{N-1} f(x_{i|k}, u_{i|k}) + f_T(x_{N|k}) \\
        \text{s.t.} ~~ &g(x_{i|k}, u_{i|k}) \leq 0, ~~~~~~~~ \forall i = \{0, \ldots, N-1\} \\
        &g_T(x_{N|k}) \leq 0, \\
        &x_{0|k} = x(k),\\
        &h(x_{i|k}, u_{i|k}) = x_{i+1|k}, ~~\forall i = \{0, \ldots, N-1\},
    \end{aligned}
\end{equation}
which computes a control input $u(k) = u_{0|k}$ for a discrete nonlinear system 
\begin{equation}\label{eq:sysDyn}
x(k+1) = h(x(k), u(k)), ~~~~ k\in \mathbb{N},
\end{equation}
where $u(k) \in\mathbb{R}^q$ and $x(k)\in\mathbb{R}^n$.
To simplify the notation in the remainder of the paper, consider
\begin{equation*}
    \begin{aligned}
        z_{i|k} &= \mat{x_{i|k}^T & u_{i|k}^T}^T, ~~~ \forall i = \{0, \ldots, N-1\} \\
        z_{N|k} &= x_{N|k}.
    \end{aligned}
\end{equation*}
With this notation, we linearize \eqref{eq:NMPC_problem} over an estimated guess trajectory $z^g_{i|k}$ \cite{SQP:Gros2016}, which results in the following QP problem:
\begin{equation}
    \label{eq:Lin_problem}
    \begin{aligned}
        \min_{\Delta z_{i|k}} ~~ &\sum_{i=0}^{N} \frac{1}{2}\Delta z_{i|k}^TQ_i\Delta z_{i|k} + \Delta z_{i|k}^TF_i \\
        \text{s.t.} ~~ &M_i\Delta z_{i|k} \leq -s_i, ~~~~~~~~~~~~~~\forall i = \{0, \ldots, N-1\} \\
        &M_N\Delta z_{N|k} \leq -s_N, \\
        &E\Delta z_{0|k} = x(k) - Ez^g_{0|k}\\
        &E\Delta z_{i+1|k} - A_i\Delta z_{i|k} = r_{i+1}, ~~\forall i = \{0, \ldots, N-2\} \\
        &\Delta z_{N|k} - A_{N-1}\Delta z_{N-1|k} = r_{N},
    \end{aligned}
\end{equation}
where the optimization variable $\Delta z_{i|k}$ is the optimal step direction with respect to $z^g_{i|k}$, $E = \mat{I_{n\times n} & \mathbf{0}_{n\times q}}$ and
\begin{equation}
    \label{eq:SQP_step}
    \begin{aligned}
        Q_i &= \left.\frac{\partial^2 f(z)}{\partial z^2}\right|_{z^g_{i|k}}, F_i = \left.\frac{\partial f(z)}{\partial z}\right|_{z^g_{i|k}}, M_i = \left.\frac{\partial g(z)}{\partial z}\right|_{z^g_{i|k}},\\
        Q_N &= \left.\frac{\partial^2 f_T(z)}{\partial z^2}\right|_{z^g_{N|k}}, F_N = \left.\frac{\partial f_T(z)}{\partial z}\right|_{z^g_{N|k}}, s_i = g(z^g_{i|k}), \\
        M_N & = \left.\frac{\partial g_T(z)}{\partial z}\right|_{z^g_{N|k}},  s_N = g_T(z^g_{N|k}),\\
        A_i &= \left.\frac{\partial h(z)}{\partial z}\right|_{z^g_{i|k}}, r_i = h(z^g_{i|k}) - Ez^g_{i+1|k},\\
        r_N &= h(z^g_{N-1|k}) - z^g_{N|k}. 
    \end{aligned}
\end{equation}
To express the cost function without the prediction time index $i$, consider the following augmented vectors and matrices:
\begin{equation}
    \label{eq:SQPnewNotation}
    \begin{aligned}
        \mathbf{z}^g &= \ColVec{z^g_{0|k}}{z^g_{N|k}}, \mathcal{A} = \mat{E & & & & \\ -A_0 & \sddots & & & \\ & \sddots & E & & \\ & & -A_{N-1} & I}, \\
        \mathcal{Q} &= \diag{Q_1}{Q_N}, \mathcal{M} = \mat{M_0 & &\\ & \ddots & \\ & & M_N}, \\
        \mathcal{F} &= \ColVec{F_1}{F_N}, \mathbf{r} = \mat{x(k)-Ez^g_{0|k} \\ r_1 \\ \vdots \\ r_{N}}, \mathbf{s} = \ColVec{s_1}{s_N},
    \end{aligned}
\end{equation}
where $\mathbf{z}^g \in\mathbb{R}^{p}$, with $p=N(n+q)+n$, and corresponding control problem
\begin{equation*}
    \begin{aligned}
        \costFcn{\Delta \mathbf{z}}{\Delta \mathbf{z}^T\mathcal{Q}\Delta \mathbf{z} + \Delta \mathbf{z}^T\mathcal{F}}{\mathcal{M}\Delta \mathbf{z}\leq -\mathbf{s} \\ & \mathcal{A}\Delta \mathbf{z} = \mathbf{r}}.
    \end{aligned}
\end{equation*}

The solution $\Delta \mathbf{z}$ of the quadratic program \eqref{eq:Lin_problem} is used to update $\mathbf{z}^g_+ = \mathbf{z}^g + \alpha \Delta \mathbf{z}$, for some $\alpha \in (0, 1]$. The step size $\alpha$ can be estimated using backtracking on the Armijo/Wolfe conditions \cite{NLMPCinSQP:Torrisi2016}, or using algebraic approximations \cite{EfficientMPCforWDN:Verheijen2022}. The final solution of the nonlinear problem is obtained by sequentially linearizing problem \eqref{eq:NMPC_problem} over the updated $\mathbf{z}^g$ until convergence. Convergence is achieved if $\mathbf{z}^g$ satisfies the KKT corresponding to \eqref{eq:NMPC_problem}, i.e.
\begin{equation}
    \label{eq:KKT_NL}
    \begin{aligned}
        0 &= \mathcal{F} + \mathcal{M}^T\lambda + \mathcal{A}^T \mu \\
        \lambda &\geq 0 \\
        0 &= \lambda \circ \mathbf{s} \\
        0 &= \mathbf{r},
    \end{aligned}
\end{equation}
which holds if $\Delta \mathbf{z}$ satisfies:
\begin{equation}
    \label{eq:KKT_step}
    \begin{aligned}
        0 &= \mathcal{Q} \Delta \mathbf{z} \\
        0 &= \mathcal{A} \Delta \mathbf{z} \\
        0 &= \lambda \circ \mathcal{M} \Delta \mathbf{z},
    \end{aligned}
\end{equation}
where $\circ$ denotes the Hadamard product. Therefore, if $\Delta \mathbf{z}$ satisfies \eqref{eq:KKT_step}, $\mathbf{z}^g$ satisfies \eqref{eq:KKT_NL}, which implies solving the last QP was not necessary (aside from validating convergence).

Assuring that the initial guess for $\mathbf{z}^g$ is close to the final solution is key to fast convergence \cite{SQP:Gros2016}. When the solution of the previous time--step is available, the initial guess of the current time--step can be simply obtained as
\begin{equation}
    \label{eq:SQPGuessShift}
    \begin{aligned}
        z^g_{i|k} &= z^g_{i+1|k-1}, ~~~~ \forall i=\{0, \ldots, N-2\} \\
        z^g_{N-1|k} &= \mat{x^g_{N|k-1} \\ u^g_{N-1|k-1}} \\
        z^g_{N|k} &= h(z^g_{N|k-1}, u^g_{N-1|k-1}).
    \end{aligned}
\end{equation}
Otherwise, some (preferably feasible) initial starting point must be estimated. Furthermore, if under any circumstances, the control problem differs significantly from the previous one (e.g., setpoint change), fast convergence is still not guaranteed. Next, we introduce a method to improve convergence, by combining parallel shooting of the initial guess and the step size with SQP.

\section{Parallel Shooting SQP}\label{sec:PSSQP}
\begin{figure*}[t!]
    \centering
    \includegraphics[width=0.95\textwidth, trim={0cm 0cm 0cm -0.5cm}]{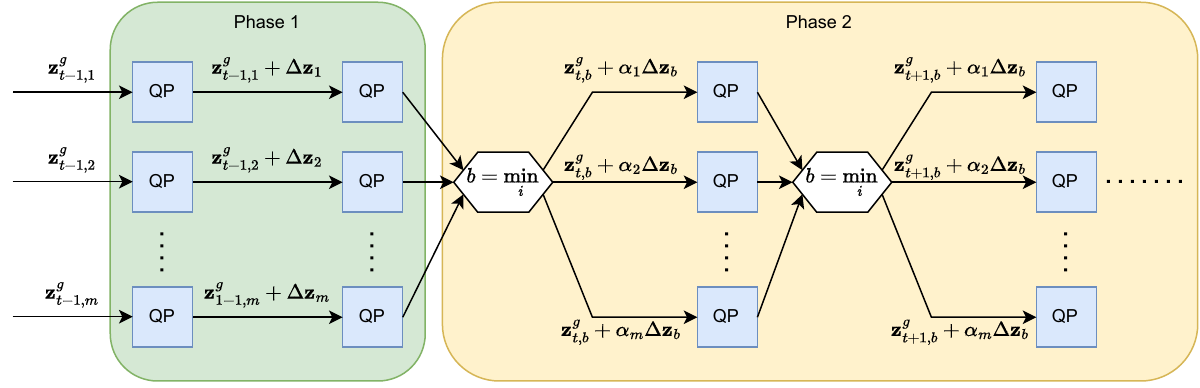}
    \caption{GPU diagram of the PS-SQP algorithm}
    \label{fig:BlockDiagram}
\end{figure*}
Provided the sequential nature of SQP, convergence requires a varying number of iterations. This generally depends on the nonlinear behaviour that is observed between the initial point and the actual optimal solution. In this section, we propose an algorithm that uses the advantages of parallel computing in SQP to reduce the amount of iterations required to converge. Herein, we propose two sequential phases: shooting over various initial trajectories and shooting over the step-size.
\begin{remark}
    To provide clarity to the reader, we only consider the entire vector of optimization variables $\bz$, and introduce the new subscript $\bz_{t,j}$, where $t$ denotes the SQP iteration and $j$ denotes the parallel process index. Since the SQP algorithm is used to solve a nonlinear MPC problem at every time instant $k$, and we focus on the solution for any such time instant, the time index $k$ is omitted.
\end{remark}
\textit{Phase 1: Parallel Shooting in multiple initial trajectories}\\
In this first phase, we construct $m$ initial guesses of the initial trajectory and generate corresponding QP problems that can be solved in parallel. This increases the likelihood that one of the initial guesses is closer to the final solution and could thus converge faster and is similar to the parallelization method proposed by \cite{GPUaccSQP:Hu2017}. Note however, a method to generate the initial trajectories for the parallel QPs is not specified in \cite{GPUaccSQP:Hu2017}, to the best of our understanding. Our proposal to generate various initial trajectories is the following
\begin{equation}
    \label{eq:PSInit}
    \begin{aligned}
        \bz^g_{0,j} = \bz^g_{0,1} + \epsilon_j, ~~ \forall \epsilon_j\in \ker\{\mathcal{A}\} \text{ and } j \in \{2, \ldots, m\}
    \end{aligned}
\end{equation}
where $\mathcal{A}$ is derived from $\bz^g_{0,1}$ as in \eqref{eq:SQP_step} and \eqref{eq:SQPnewNotation}. This can be reasoned through the following: any trajectory $\bz_{t,j}$ is a trajectory of the nonlinear system \eqref{eq:sysDyn} if the corresponding $\mathbf{r}_{t,j} = 0$, with $\mathbf{r}_{t,j}$ computed as in \eqref{eq:SQP_step} and \eqref{eq:SQPnewNotation}. A suitable initial trajectory for the SQP problem is a trajectory with minimal linearization error (as this trajectory is then close the system dynamics). For any offset $\epsilon_j$ from \eqref{eq:PSInit}, the linearization error can be expressed as:
\begin{equation}
    \label{eq:InitTrajErr}
    \begin{aligned}
        \mathbf{r}_{0,j} &= \mathbf{r}_{0,1} + \mathcal{A}\epsilon_j + \mathcal{O}(\|\epsilon_j\|^2) = \mathcal{O}(\|\epsilon_j\|^2).
    \end{aligned}
\end{equation}
This is a valid assumption, as the solution of the last time iteration satisfies the system dynamics and the shifted estimate \eqref{eq:SQPGuessShift} is computed such that it also satisfies system dynamics \eqref{eq:sysDyn}. Then, by keeping $\|\epsilon_j\|^2$ appropriately bounded, it can be assumed that the corresponding SQP problem is feasible.

For practical use, note that if
\begin{equation}\label{eq:PractInitVec}
    \begin{aligned}
        \epsilon_j = (I - \mathcal{A}^\dagger\mathcal{A})\omega, \text{ then } \epsilon_j\in \ker\{\mathcal{A}\} ~~ \forall \omega\in \mathbb{R}^{p}.
    \end{aligned}
\end{equation}
Thus, one can randomly generate any $\omega$ to obtain a suitable $\epsilon$. Here, $(\cdot)^\dagger$ denotes the generalized pseudo--inverse. It should be mentioned that computing pseudo--inverses can be computationally expensive. Alternatively, it is also possible to generate a random input offset, and compose $\bz^g_{0,j}$ by iterating over the discretized system dynamics. 

By iterating over this set of parallel trajectories, either one (or more) will satisfy the convergence conditions, or all will get stuck in a cycle. To prevent the latter, we introduce phase 2.
\begin{remark}
    Note that the proposed method to generate guess trajectories in \eqref{eq:PSInit} is not the only suitable method. Technically, $\mathbf{z}^g_{0,j}$ can be initialized with any arbitrary value, as long as the corresponding QP is feasible. 
\end{remark}


\textit{Phase 2: Parallel Shooting for multiple step-sizes}\\
To trigger this phase, one of the following conditions must be satisfied, where
\begin{equation}\label{eq:LinErr}
    \begin{aligned}
        \mathbf{e}_{t,j} = \norm{\mat{\mathcal{Q}\Delta \mathbf{z}_{t,j} \\ \lambda_j \circ \mathbf{s}_{t,j} \\ \gamma\mathbf{r}_{t,j}}}
    \end{aligned}
\end{equation} denotes the residual linearization error for process $j$ at iteration $t$ and $\mathbb{I} = {1, \ldots, m}$:
\begin{subequations}
\begin{enumerate}
    \item no convergent trajectories, i.e. (given $t \geq 1$) \begin{equation}\label{eq:P2cond1}\max_{j\in\mathbb{I}}(\mathbf{e}_{t,j}-\mathbf{e}_{t-1,j})>0,\end{equation}
    \item all trajectories are close to equal, i.e. \begin{equation}\label{eq:P2cond2}\mathbf{z}_{t,1}\approx \mathbf{z}_{t,2} \approx \ldots \approx \mathbf{z}_{t,m},\end{equation}
    \item all trajectories are stuck in a loop, i.e. \begin{equation}\label{eq:P2cond3}\Delta \mathbf{z}_{t-1,j} = -\Delta \mathbf{z}_{t,j}, \forall j\in\mathbb{I}.\end{equation}
\end{enumerate}
\end{subequations}
In this phase, we linearize every parallel NLP over the best trajectory of the previous iteration, but now give each a different step-size $\alpha$ for the step $\Delta \mathbf{z}$. If the phase is triggered, the algorithm will not return back to phase 1 until convergence is achieved. The distribution of the different step-sizes to the parallel QPs is open to different designs. Using $\alpha_j = j/m$ already provided good results, but other strategies could be explored. Nonetheless, it is recommended to always keep $\alpha=1$ in the set of step-size shooting.

\begin{algorithm}[H]
    \caption{Parallel Shooting SQP}\label{alg:sqp}
    \begin{algorithmic}[1]
    \Require $\mathbf{z}^g_{0,1}$, $f(z)$, $g(z)$, $h(z)$, $m$, $\delta$, $\gamma$
    \For{$j = 2:m$}
        \State Generate vector $\epsilon_j$ as in \eqref{eq:PractInitVec}
        \State $\mathbf{z}^g_{0,j} = \mathbf{z}^g_{0,1} + \epsilon_j$
    \EndFor
    \State $t \gets 0$
    \State $P2 \gets 0$
    \State $e_{-1,j} \gets 2\delta, ~\forall j\in\mathbb{I}$
    \While{$\min_{j\in\mathbb{I}}(\mathbf{e}_{t-1,j})\geq \delta$}
        \ParFor{$j=1:m$}
        \State Build Linearized problem \eqref{eq:Lin_problem} using $\mathbf{z}^g_{t,j}$
        \State Solve QP and obtain $\Delta \mathbf{z}_{t,j}$, $\lambda_j$
        \State Compute $\mathbf{e}_{t,j}$ as in \eqref{eq:LinErr}
        \EndFor
        \If{\eqref{eq:P2cond1} \textbf{ or } \eqref{eq:P2cond2} \textbf{ or } \eqref{eq:P2cond3}}
            \State $P2 \gets 1$
        \EndIf
        \If{$P2 = 0$}
            \State $\mathbf{z}^g_{t+1, j} \gets \mathbf{z}^g_{t,j} + \Delta \mathbf{z}_{t,j}, \forall j\in\mathbb{I}$ 
        \Else
            \State $b \gets \argmin_{j\in\mathbb{I}}(\mathbf{e}_{t,j})$
            \For{$j = 1:m$}
                \State $\alpha_j \gets \frac{j}{m}$
                \State $\mathbf{z}^g_{t+1, j} \gets \mathbf{z}^g_{t,b} + \alpha_j\Delta \mathbf{z}_{t,b}$
            \EndFor
        \EndIf
        \State $t \gets t+1$
    \EndWhile
    \State $b \gets \argmin_{j\in\mathbb{I}}(\mathbf{e}_{t-1,j})$
    \Ensure $\mathbf{z}^g_{t+1,b}$
    \end{algorithmic}
\end{algorithm}
When convergence of the Parallel Shooting Sequential Quadratic Programming (PS-SQP) algorithm is achieved, the first control input can be applied to the system following the receding horizon principle. The complete PS-SQP solver for a single sampling interval is explained in Algorithm \ref{alg:sqp}, where the parallel part is described, as well as the implementation of the two phases. The data communication between the QPs in sequential steps is illustrated in Figure \ref{fig:BlockDiagram}. Mind that this illustrates just 4 sequential steps, whereas in practice, both parallelization techniques might take more sequential steps for convergence (similarly, it could also finish in just 2 or even 1 sequential step).

As shown in Algorithm \ref{alg:sqp}, convergence is achieved if the linearization error $\mathbf{e}_{t,j} < \delta$. Selecting a suitable $\delta$ can require some testing and is highly dependent on the weights in $\mathcal{Q}$. Furthermore, assuming large weights in $\mathcal{Q}$, this term will be dominant in the assessment of the error \eqref{eq:LinErr}. We thus also introduce $\gamma$, which can be used to boost the influence of the error in the equality constraints. Furthermore, if the PS-SQP algorithm switches to phase two, a properly tuned $\gamma$ will prioritize solutions with smaller norms of $\mathbf{r}_t$ first.
\begin{remark}
    As an alternative to Algorithm \ref{alg:sqp}, one can initialize for each core in Phase 1 multiple parallel cores that each test different values for $\alpha$. This does require a significant larger amount of available parallel cores. However, this could improve convergence speed.
\end{remark}

\subsection{GPU implementation of PS-SQP}
To enable the full potential of the PS-SQP algorithm it can be implemented on a GPU. By doing so, there are two main advantages to consider depending on the level of GPU implementation: First, the QPs can be computed in parallel in $m$ GPU kernels (see Figure \ref{fig:BlockDiagram}). However, one main design challenge of such parallel implementation is to deal with large overhead due to communication over the interconnect between the GPU kernels. Since the transfer of large matrices is required between the iterations, the communication overhead may be high. To address this challenge, the large matrices in the linearized control problem can be transferred on the GPU kernels in a sparse format. Furthermore, since the linearized problem matrices do not change the sparsity pattern, regardless of the operating point, every GPU kernel can store the same sparsity pattern once and then only needs to receive the individual values from the CPU. This strategy can massively reduce the communication overhead, which is considered a significant bottleneck in GPU--accelerated computing \cite{Data_transfer}. The second advantage is that every individual QP can be computed in parallel on the GPU as well. Here, available methods such as qpDUNES \cite{qpDUNES:Frasch2015}, accelerated ADMM \cite{GPUADMM:Schubiger2020}, parallel Interior Point \cite{GPUaccSQP:Hu2017} or PQP \cite{PQP:Brand2011} can provide a significant speedup. However, each QP will utilize a block of GPU kernels, which reduces the number of kernels available for the PS-SQP.

\section{Illustrative Examples}\label{sec:Examples}
To illustrate the effectiveness of the algorithm, we implemented the solver on an MPC problem for an inverted pendulum system as shown in \cite{InvertedPendulumModel:Prasad2014}. This system has strong nonlinear dynamics, and the solver is prone to get stuck at challenging setpoints \cite{SQPConvergence:Torrisi2018}. Additionally, the PS-SQP solver is also tested on a small water distribution network. These systems generate large--scale NMPC problems.

\subsection{Inverted Pendulum}\label{ssec:InvPendulum}
Consider the following continuous time dynamics \cite{InvertedPendulumModel:Prasad2014}:
\begin{equation}
    \label{eq:InvPendulum}
    \begin{aligned}
        \dot{x}_1 &= x_2 \\
        \dot{x}_2 &= \frac{u \cos{x_1} - (M+m)g\sin{x_1} + ml(\cos{x_1}\sin{x_1})x_2^2}{ml\cos^2{x_1}-(M+m)l} \\
        \dot{x}_3 &= x_4 \\
        \dot{x}_4 &= \frac{u+ml\sin{x_1}x_2^2 - mg\cos{x_1}sin{x_1}}{M+m-m\cos^2{x_1}}
    \end{aligned}
\end{equation}
where $x_1$ is the pendulum angle, with $x_1 = 0$ corresponding to the upright position, $x_2$ is the angular velocity, $x_3$ is the position of the cart, $x_4$ is the cart velocity. The remaining parameters are selected as \cite{InvertedPendulumModel:Prasad2014}: 
\begin{itemize}
    \item mass of cart $M=2.4kg$;
    \item mass of pendulum $m=0.23kg$;
    \item length of pendulum $l=0.36m$;
    \item gravity constant $g=9.81m/s^2$.
\end{itemize} 
The model in \eqref{eq:InvPendulum} is discretized using backwards Euler with a sampling period of $T_s = 0.02s$. However, in the simulation, the system response is computed using an \texttt{ode45} solver for the continuous time dynamics. The position of the cart is constrained by $-10 \leq x_1 \leq 10$, and the input force is constrained to $-500 \leq u \leq 500$. The controller is tasked to track a position reference with amplitude $\pm 3 [m]$ for the cart, whilst keeping the pendulum upright. For this, we construct a standard quadratic cost function over a prediction horizon of $N=40$, with weights $Q=\text{diag}(100, 0.1, 500, 0.1)$, $R=0.001$ and $Q_T =\text{diag}(1000, 10, 500, 10)$. We assume that the full state is measurable. The pendulum starts initially in the downwards position. 

\begin{figure}[h!]
    \centering
    \includegraphics[width=0.45\textwidth, trim={2cm 0.1cm 2cm 0.5cm}]{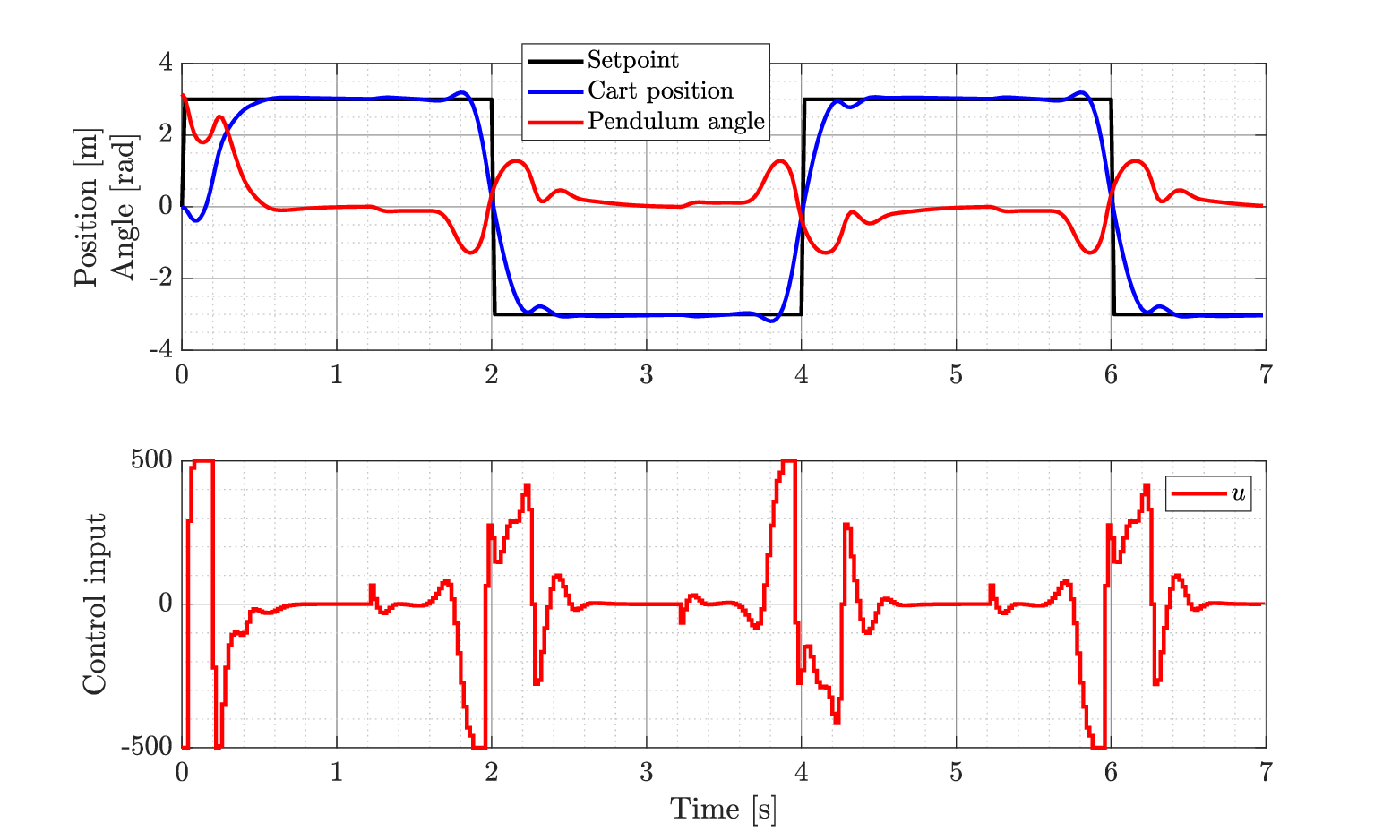}
    \caption{Simulation results of the inverted pendulum}
    \label{fig:SimPlot}
\end{figure}

\begin{figure}[h!]
    \centering
    \includegraphics[width=0.45\textwidth, trim={2cm 0.1cm 2cm 0.5cm}]{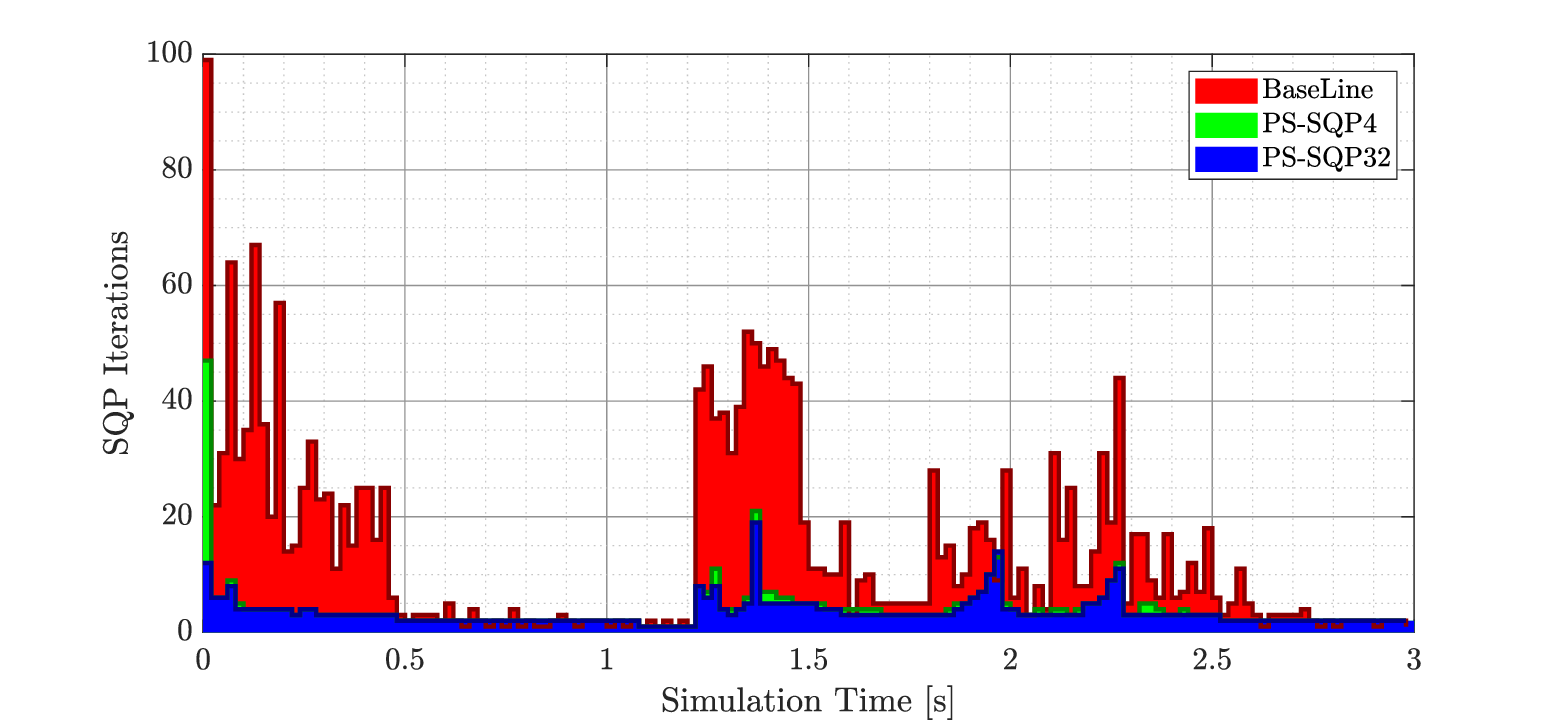}
    \caption{Number of SQP iterations required per simulation time sample}
    \label{fig:SQPComparison}
\end{figure}

Figure \ref{fig:SimPlot} illustrates the simulated response. The steep setpoint change caused the pendulum to take a significant swing, which shows the effectiveness of the nonlinear control law. 
To provide a baseline, we constructed the same control problem to be solved by MATLAB's \texttt{fmincon.m} using the SQP option. Note that to accelerate the convergence of this routine, the gradients of the nonlinear dynamics are provided. For both the baseline simulation and the implemented PS-SQP simulations, we used $\delta = 0.5$ as the linearization tolerance. The comparison in SQP iteration between the baseline and the PS-SQP algorithm is illustrated in Figure \ref{fig:SQPComparison}. Note that the x-axis represent the simulation time in Figure \ref{fig:SimPlot}, but only up to the first 3 seconds. When the nonlinear dynamics must be excited by the controller, a noticeable reduction can be observed. Furthermore, as also illustrated, a small amount of parallel cores is already sufficient for this NLP, as the improvement between 32 cores and 4 cores is minor. Although hard to observe, the BaseLine does converge in less steps for a few time--samples, most notably when the system is almost static.

\subsection{Small Water Distribution Network}\label{ssec:simWDN}
\begin{figure}[h]
\vspace{0.2cm}
	\centering
	\resizebox{0.98\hsize}{!}{%
	\begin{tikzpicture}
		\draw (0, 0) pic{tank={H1, $h_1$}};
		\draw (9.5, 4) pic{tank={H6, $h_6$}};
		\draw (13.5, 0) pic{tank={H5, }};
		\draw (2, 0) pic{pump={P1, $u_1$}};
		\draw (10, 1.5) pic{pumpVert={P2, $u_2$}};
		\node (J2) at (10,0) {};
		\node (J4) at (12,4) {};
		\node (J3) at (14,4) {};
		\draw[ultra thick] (H1_right) -- (P1_in);
		\draw[ultra thick] (P1_out) -- (J2);
		\draw[ultra thick] (J2) -- (H5_left);
		\draw[ultra thick] (J2) -- (P2_in);
		\draw[ultra thick] (P2_out) -- (H6_bottom);
		\draw[ultra thick] (H6_right) -- (J4);
		\draw[ultra thick] (J4) -- (J3);
		\draw[ultra thick] (J3) -- (H5_top);
		\draw[ultra thick, fill=black] (J2) circle [radius = 0.1];
		\draw[ultra thick, fill=black] (J3) circle [radius = 0.1];
		\draw[ultra thick, fill=black] (J4) circle [radius = 0.1];
		\node () at ([yshift=-0.4cm]J2) {$h_2$};
		\node () at ([xshift=0.4cm]J3) {$h_3$};
		\node () at ([yshift=-0.4cm]J4) {$h_4$};
		\node () at ([xshift=0.4cm]H5_right) {$h_5$};
		\draw[->, ultra thick] (J3) -- (14, 5);
		\draw[->, ultra thick] (J4) -- (12, 5);
		\node () at (12.5,4.5) {$d_1$};
		\node () at (14.5,4.5) {$d_2$};
	\end{tikzpicture}}
	\caption{Water Network Example}
	\label{fig:AdaptedNetwork}
\end{figure}
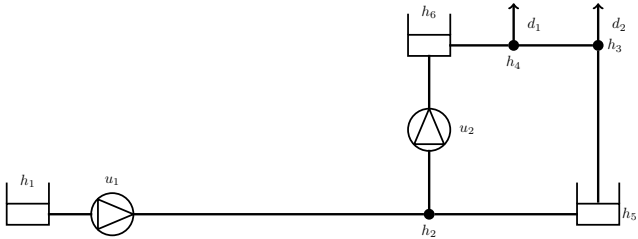

In this example we consider a small water distribution network, as seen in Figure \ref{fig:AdaptedNetwork}. This water network can be modeled using the balance of mass and energy equations \cite{EfficientMPCforWDN:Verheijen2022}, i.e. 
\begin{equation}\label{eq:SysEqsWDN}
	\begin{aligned}
		\mathbf{h}(k+1) &= \mathbf{h}(k) + A_q\mathbf{q}(k) + B_q\mathbf{u}(k) \\
		\mathbf{0} &= G_q\mathbf{q}(k) + G_u\mathbf{u}(k) + G_d\mathbf{d}(k)\\
		\mathbf{0} &= F_h\mathbf{h}(k) + \Phi(\mathbf{q}(k)),
	\end{aligned}
\end{equation}
where it can be noted that the only nonlinearities are contained in $\Phi(\mathbf{q}(k))$, which are the pressure losses in the pipes. 
For the definition of matrices in \eqref{eq:SysEqsWDN}, the control problem and constraints, please consider the results in \cite{EfficientMPCforWDN:Verheijen2022}.
\begin{remark}
    Compared to \cite{EfficientMPCforWDN:Verheijen2022}, the pump cost is implemented as 
    \begin{equation}\label{eq:CostPumps}
        \begin{aligned}
            l_3(k+i) &= \left(\frac{\rho g\varepsilon(k+i)}{850}\right)^2 \times \\ & ~~~\sum_{p} \left((\Delta h^p_{i|k})^2+3(\Delta h^p_{\text{max}})^2\right)(u^p_{i|k})^2
        \end{aligned}
    \end{equation}
    where $u^p$ is the control flow through pump $p$, $\Delta h^p$ the corresponding head gain, $\varepsilon(t)$ the pump tariff, $\Delta h^p_{\text{max}}$ is the maximum head the pump can assert, $\rho$ and $g$ are physical constants for the density and gravity. The cost in \eqref{eq:CostPumps} is then correctly linearized to build the QPs. The additional term $\Delta h^p_{\text{max}}$ is used to ensure \eqref{eq:CostPumps} has a positive semi-definite Hessian.
\end{remark}

\begin{figure}[h!]
    \centering
    \includegraphics[width=0.45\textwidth, trim={2cm 0.1cm 2cm 0.5cm}]{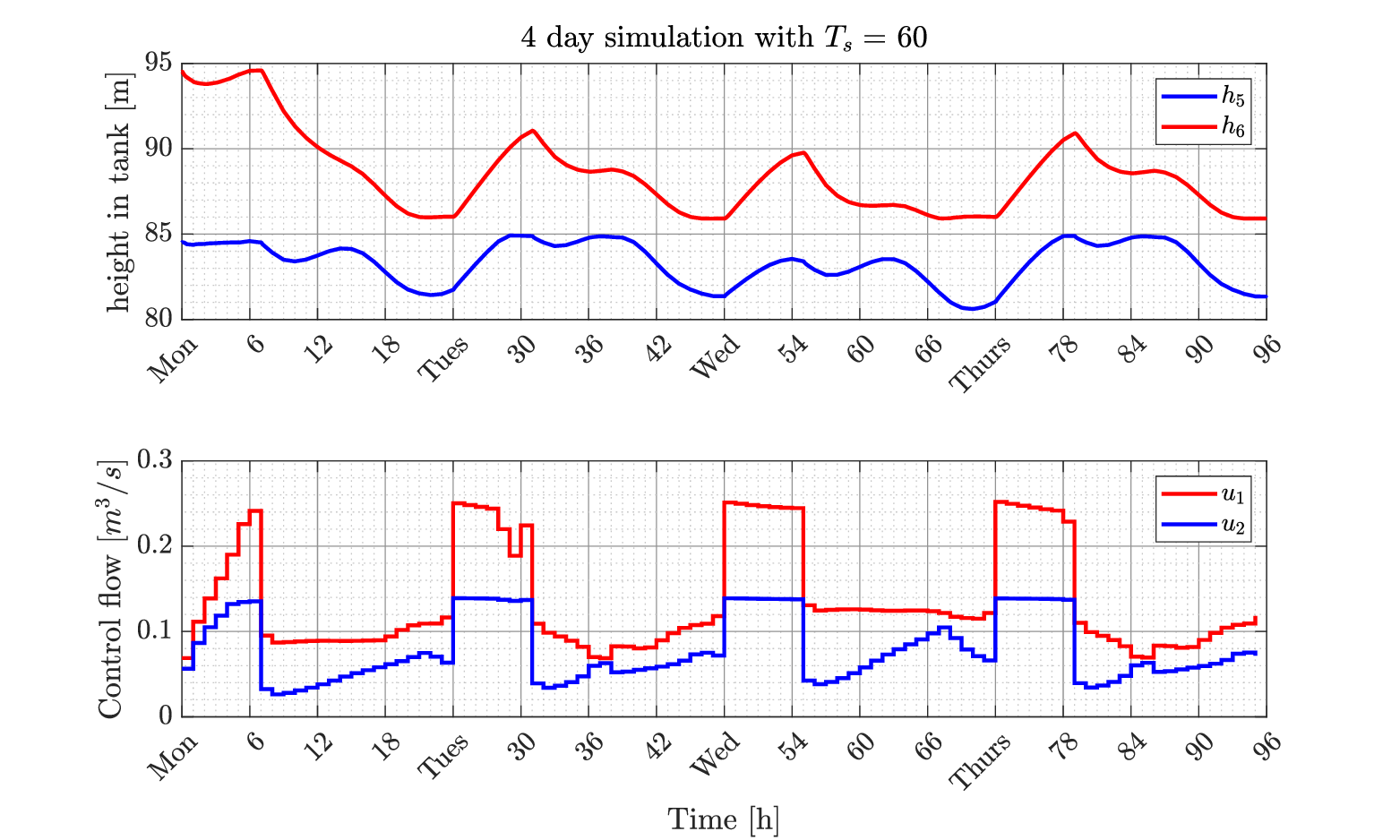}
    \caption{Simulation results of the small water network}
    \label{fig:SimulationWDN}
\end{figure}

The corresponding nonlinear control problem is, compared to the inverted pendulum, not only nonlinear in the equality constraints, but also in the cost function and the inequality constraints. See Figure \ref{fig:SimulationWDN} for the simulation results of the water network, controlling the flow of the two pumps, while keeping the water levels in the tanks at a satisfactory level. Hereby, the demand patterns are slightly different than the predicted demand, following the mismatch used in \cite{EfficientMPCforWDN:Verheijen2022}. Furthermore, we considered a sampling period of $T_s = 1 [h]$, a prediction horizon of $N=24$ (an entire day) and the linearization tolerance has been set to $\delta = 0.0001$.

\begin{figure}[h!]
    \centering
    \includegraphics[width=0.45\textwidth, trim={2cm 0.1cm 2cm 0.5cm}]{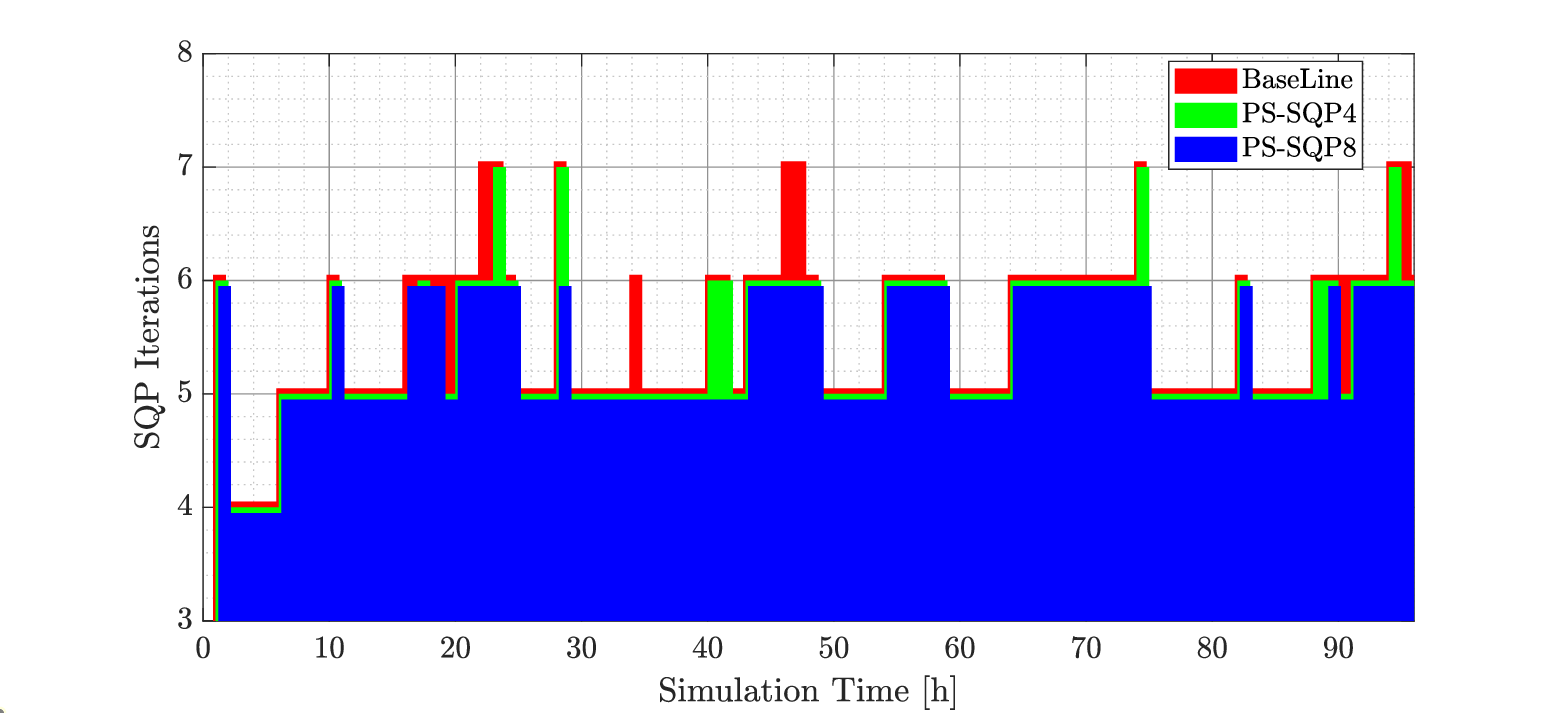}
    \caption{Number of SQP iterations required per simulation time sample}
    \label{fig:SQPComparisonWDN}
\end{figure}

In this example, we compared the standard, nonparallel SQP algorithm as in \cite{SQP:Gros2016} with the PS-SQP with either 4 cores or 8 cores. Following the results in Figure \ref{fig:SQPComparisonWDN}, we observe that minor improvements are made compared to the BaseLine. 
This can be a result of having a low linearization error between the shifted optimal solution of the previous iteration and the current optimal solution. For this water network problem, the nonlinear components, namely the pipe friction coefficients and the pump cost/constraints, are almost linear (enough to often control them considering a fully/piece--wise linearized model \cite{UrbanWDN:Leirens2010}\cite{OptDist:Verleye2013}). In this case, the respective optimal Newton step size $\alpha = 1$ guarantees fastest convergence, see \cite[Thrm 3.5]{NumOptBook:Nocedal}. This makes shooting in $\alpha$ unnecessary, and thanks to Phase 1, rarely employed (as all trajectories are sufficiently decreasing). Therefore, we only obtain a faster convergence if any of our guessed trajectories achieves a cost less than the bound, while the default trajectory (i.e., the trajectory that is initialized as the shifted previous optimal trajectory) slightly exceeded the bound. 


\section{Conclusions}\label{sec:Conclusion}
In this paper, we have proposed a parallel algorithm to solving nonlinear model predictive control problems using sequential quadratic programming. This algorithm is built using a a two-phase approach that first assesses sequential convergence over many initial trajectories in parallel. However, if none converged, the algorithm starts varying the Newton step-size in parallel instead. Through this parallel shooting approach, it was expected that the number of iterations to converge to the nonlinear solution can be decreased. Furthermore, the algorithm can be further expanded and accelerated by implementing it on GPUs. We illustrated the effectiveness of the proposed PS-SQP in some benchmark examples for nonlinear model predictive control. The developed PS-SQP parallel solver converged faster when significant nonlinear behaviour is excited in the NMPC horizon.
\\
\bibliographystyle{IEEEtran}
\bibliography{references}
\end{document}